\documentclass[preprint,showpacs,prb,superscriptaddress]{revtex4}

\usepackage{graphicx}

\def\bea{\begin{eqnarray}}
\def\eea{\end{eqnarray}}

\def\a{\alpha}

\def\p{\partial} 
\def\nn{\nonumber}

\def\la{\langle}
\def\ra{\rangle}

\def\n{\eta}
\def\g{\gamma}

\def\f{\frac}

\def\Hh{\hat{H}}
\def\Gh{\hat{G}}

\def\gh{\hat{\g}}

\def\T{{\bf{T}}}
\begin{document}

\title{ Classical limit of master equation for harmonic oscillator
  coupled to oscillator bath with separable initial conditions }
\author{Subhashish Banerjee}
\email{subhashishb@rri.res.in}
\author{Abhishek Dhar}
\email{dabhi@rri.res.in}
\affiliation{Raman Research Institute, Bangalore 560080}
\date{\today} 

\begin{abstract}
The  equation for the Wigner function describing the reduced
dynamics of a single harmonic oscillator,
coupled to an oscillator bath, was obtained by Karrlein and Grabert
[Phys. Rev. E {\bf 55}, 153 (1997)]. It was shown that for
some special correlated initial conditions the  equation reduces, 
in the classical limit, to the
corresponding classical Fokker-Planck equation obtained by Adelman
[J. Chem Phys. {\bf 64}, 124 (1976)]. However for separable
initial conditions the Adelman equations were not recovered. We
resolve this problem by showing that, for separable initial
conditions, the classical Langevin equation obtained from the
oscillator bath model is somewhat different
from the one considered by Adelman. We obtain the corresponding
Fokker-Planck equation and show that it exactly matches  the
classical limit of the equation for the  Wigner function obtained
from the master equation for separable 
initial conditions. We also discuss why the special correlated initial 
conditions correspond to Adelman's solution.
\end{abstract}

\maketitle

\section{Introduction}
The concept of `open' quantum systems is a ubiquitous one in that any 
real system is never fully isolated. It is usually in contact with 
a larger system, constituting its environment, which  effects 
its dynamics. 
Ford, Kac and Mazur \cite{fkm65} suggested the first microscopic model
describing dissipative effects. They considered a system which was coupled to a bath 
modelled by a collection of an infinite number of harmonic oscillators. 
Interest in
quantum dissipation was intensified by the work of Caldeira and
Leggett \cite{cl83} who used the influence functional approach,
developed by Feynman and Vernon \cite{fv63}, to discuss quantum
Brownian motion. They considered the case where the system and its environment
were initially uncorrelated, the so called separable initial condition. 
This was generalized to the situation where
initial correlations exist between the system and its environment by
Hakim and Ambegaokar \cite{hakim}, Smith and Caldeira \cite{sc87},
Grabert, Schramm and Ingold \cite{grabert}, Chen, Lebowitz and Liverani
\cite{cll89} and Banerjee and Ghosh \cite{bg03} among others.
The master equation for the quantum Brownian motion of a harmonic
oscillator in a bosonic bath of harmonic oscillators was obtained
by Haake and Reibold \cite{hr85} and later by Hu, Paz and Zhang
\cite{hpz92} who used path integral methods in their derivation.
A derivation of the Wigner transform of this equation was also
given by Halliwell and Yu \cite{hy96} and a solution of this
Fokker-Planck equation with time dependent coefficients was given by
Ford and O'Connell \cite{ford}.  

In a very interesting paper Karrlein and Grabert \cite{karrlein}
studied the master equation for the reduced density matrix of
the system for generalized correlated initial conditions. They showed that
in general it is not possible
to obtain a Liouvillean operator for the master equation of the reduced
dynamics of the system under the influence of its environment. However 
for some specialized correlated initial conditions like the one discussed
by Hakim and Ambegaokar \cite{hakim}, called the thermal initial
condition, and for the case of the time evolution of equilibrium
correlation functions \cite{grabert}, it is possible to obtain the
Liouvillean to describe the reduced system dynamics. In both these cases 
Karrlein and Grabert \cite{karrlein} obtained the Fokker-Planck equation 
from the Wigner
transform of the master equation. They found that in the classical
limit, these equations reduced exactly to the corresponding equation
obtained by Adelman \cite{adelman} for the dissipative dynamics of a
classical harmonic oscillator driven by a Gaussian non-Markovian noise.
They then considered the case of separable initial conditions and found
that in the classical limit the corresponding Fokker-Planck equation
did not reduce to the Adelman equation. We find this a very intriguing
as well as an interesting point and the main purpose of this paper is
to look at this point more closely. This is also important to
understand in view of the fact that factorizing initial conditions are
known to have some problematic features \cite{grabert,cll89}
associated with the sudden switching on of the coupling.

The plan of this paper is as follows. In Sec.~II we obtain the Langevin 
equation for the system consisting of a harmonic oscillator coupled to a
bath of harmonic oscillators with which it is initially uncorrelated. We
indicate that our treatment is valid in the classical as well as the quantum 
regime. In Sec.~III we obtain the Fokker-Planck equation from the Langevin
equation derived in Sec.~II in the classical regime. When we compare our
equation with the one used by Adelman \cite{adelman} we find that our equation
has an additional term. This additional term is the crux which enables us
to resolve the issue. The Fokker-Planck equation obtained from the Langevin 
equation with this additional term is the exact classical limit of the corresponding
equation in the quantum regime obtained by Karrlein and Grabert \cite{karrlein}.
In Sec.~IV we discuss why the Langevin equation used by Adelman would be the
correct classical equation describing a correlated initial state.
In Sec.~V we discuss our conclusions.

\section{Langevin equation}
We consider a single harmonic oscillator described by the phase space
variables $(x,p)$ and with a natural frequncy $\omega_0$. The oscillator
is coupled to a heat bath modelled 
by $N$ independent oscillators  described by $\{X_\a,P_\a \}$ and
which have frequencies $\{ \omega_\a \}$ . To generate dissipation one eventually 
takes the limit $N\to \infty$. The full Hamiltonian is taken to
be:
\bea
H= \f{p^2}{2} + \f{ \omega_0^2 x^2}{2} +\sum_{\a=1}^N[~ \f{P_\a^2}{2
}+\f{ \omega_\a^2 X_\a^2}{2}~] - \sum_{\a=1}^N c_\a x X_\a~.  
\label{ham}
\eea
The last term represents the coupling between the system and the
bath. Introducing the shifted frequency $ \omega^2=  \omega_0^2 - \sum_\a
{c_\a^2}/{\omega_\a^2}$, we can rewrite Eq.~(\ref{ham}) in the
following equivalent form 
\bea
H =  \f{p^2}{2} + \f{ \omega^2 x^2}{2} +\sum_{\a=1}^N [~\f{P_\a^2}{2
  }+\f{\omega_\a^2 (X_\a- \f{c_\a x}{ \omega_\a^2}) ^2}{2}~]~. \label{eqham}
\eea
Let us now write the equations of motion for the system and bath
variables. These are given by:
\bea
\ddot{x} &=& - \omega_0^2 x + \sum_\a c_\a X_\a~,  \label{eqms}\\
\ddot{X}_\a &=& - \omega_\a^2 X_\a + c_\a x~.  \label{eqmr}
\eea
The solution of the equations of motion of the reservoir variables
gives
\bea
X_\a(t)= \cos (\omega_\a t) X_\a(0) +\f{\sin (\omega_\a t)}{\omega_\a}
\dot{X}_\a(0) +\int_0^t dt' \f{\sin \omega_\a (t-t')}{ \omega_\a} c_\a x(t')~.
\eea
Plugging these into the system's equation of motion Eq.~(\ref{eqms})
we get the following Langevin equations of motion:
\bea
\ddot{x} &=& - \omega_0^2 x + \int_0^t dt' \f{\p \g (t-t')}{\p t'} x(t') + \n(t)~, \label{lang}\\
{\rm where}~~~\g(t) &=& \sum_\a \f{c_\a^2}{ \omega_\a^2} \cos (\omega_\a t)~,\label{gameq} \\
\n(t)&=& \sum_\a c_\a [\cos (\omega_\a t) X_\a(0) +\f{\sin (\omega_\a t)}{\omega_\a}
\dot{X}_\a(0)]~. \label{noise}
\eea
Note that the discussion till now is valid both clasically and
quantum mechanically. In the quantum case the variables correspond to
operators in the Heisenberg representation. 
We now use the fact that at $t=0$ the separable initial conditions
imply that the reservoir (isolated) is initially in equilibrium and
hence we know the statistical properties of the bath variables
$\{X_\a(0),~\dot{X}_\a(0)\}$. In the quantum case we get
\bea
&&\la~\omega^2_\a  X_\a^2(0) ~ \ra =\la~ \dot{X}_\a^2(0)~ \ra = \f{\hbar \omega_\a}{2}
\coth \f{\hbar \omega_\a}{2 k_B T}~, \nn \\
&& \la~ [~X_\a(0) \dot{X}_\a(0)  + \dot{X}_\a(0) {X}_\a(0)~]~ \ra = 0. \label{bathcor} 
\eea
In the classical case we get, using equipartition, $\la \omega_\a^2 X_\a^2 \ra = \la
\dot{X}_\a^2 \ra = k_B T$ and $\la X_\a \dot{X}_\a \ra =0$  and these
also  follow from Eq.~(\ref{bathcor}) if we take the classical limit
$\hbar \omega_\a/(k_B T) \to 0$. From these properties of the bath one can
work out the correlations of the noise $\n(t)$ in
Eq.~(\ref{noise}). We get (for $t,t' > 0$) 
\bea
\la~ \n (t)~ \ra &=& 0~, \nn \\
\f{1}{2}  \la~ \n (t) \n(t')  + \n (t') \n(t)~ \ra &=& \hbar
K'(t-t')  \label{qnoise}~, \\
{\rm where}~~~ K'(t)&=& \sum_\a \f{c_\a^2}{2 \omega_\a } \cos{\omega_\a t }
\coth{ \f{\hbar \omega_\a}{2 k_B T}}~\label{kpeq}.  
\eea
Here we have used the notation in Ref.~[\onlinecite{grabert, karrlein}].
The classical limit of Eq.~(\ref{qnoise}) gives the usual fluctuation-dissipation relation
\bea
 \la \n (t) \n(t') \ra = k_B T \g(t-t') \label{cnoise}~.
\eea 
\section{Fokker-Planck equation for the classical Langevin equation} 
In this  section we focus on the classical Langevin equation which is
described by Eqs.~(\ref{lang}), (\ref{cnoise}). First we note that, after
a partial integration, we can rewrite Eq.~(\ref{lang}) in the
following form:
\bea
 \ddot{x} &=& -\omega^2 x -\g(t) x(0) - \int_0^t dt' \g (t-t') \dot{x}(t') +
\n(t)~. \label{eqlang} 
\eea
This differs from the equation considered by Adelman \cite{adelman} by the initial
condition dependent term $-\g(t) x(0)$,
the so called initial slip term \cite{gi02, ph97}
 which, as we will see, is
crucial in getting the correct Fokker-Planck equation. 
Starting from the above Langevin equation or its equivalent form
Eq.~(\ref{lang}) we get, following Adelman, the
corresponding Fokker-Planck equation.

The general solution of Eq.~(\ref{lang}) is given by:
\bea
x(t)=H(t) x(0) + G(t) p(0) + \int_0^t dt' G(t-t') \n (t')
\label{solx}~,
\eea
where $H(t)$ and $G(t)$ are solutions of the homogeneous part of
Eq.~(\ref{lang}) for the initial conditions $H(0)=1,~\dot{H}(0)=0$ and
$G(0)=0,~\dot{G}(0)=1$ respectively, and $p(t)=\dot{x}(t)$. Taking a Laplace transform of the
homogeneous equation we immediately get:
\bea
\Gh(z)&=&\f{1}{ z^2+ \omega^2+ z~\gh(z)}~, \\
\Hh(z)&=& \f{ z}{ z^2+ \omega^2+z~ \gh(z)}~, 
\eea
where $\Gh(z),~\Hh(z)$ and $\gh(z)$ are the Laplace transforms of
$G(t),~H(t)$ and $\g(t)$ respectively. 
We note that $G(t)$ is the  same Green's function, denoted by $\chi_u(t)$, in
Adelman's  \cite{adelman} paper. However $H(t)$ is different from
Adelman's $\chi_x$, which we will here denote by $F(t)$, and which
is given by
\bea
F(t)=1-\omega^2 \int_0^t dt' G(t')~.\label{feq}
\eea
From the form of the Laplace transforms $\Gh,~\Hh$ and from
Eq.~(\ref{feq}) it is also easy to get the relations
\bea
\dot{G}(t)~=~H(t) ~~~{\rm and}~~~ \dot{F}(t)~=~-\omega^2 G(t)~.
\eea
From the general solution for the position we get for the momentum 
\bea
p(t)=\dot{H}(t) x(0) +\dot{G}(t) p(0)+ \int_0^t dt'
\dot{G}(t-t')\n(t')~.
\eea
Let us use the notation $\{x_1,~x_2\}=\{x,~p\}$. The
fluctuations about the mean position and momentum are given by:
\bea
y_1(t)&=& x_1(t)- [H(t) x(0) + G(t) p(0)] = \int_0^t dt'
G(t-t')\n(t')~, \nn \\
y_2(t)&=&x_2(t)-[\dot{H}(t) x(0) +\dot{G}(t) p(0)]= \int_0^t dt'
\dot{G}(t-t')\n(t')~.\label{yeq}
\eea
In the classical case, since $\n (t)$ depends linearly on 
$X_\a(0)$ and $\dot{X}_\a(0)$ which have a 
Gaussian distribution, it follows that $\n (t)$ is also Gaussian.
Hence the distribution of 
$\{y_1,y_2\}$  at any time will also be Gaussian and is completely 
specified by the following variances, obtained from Eqs.~(\ref{cnoise}), 
(\ref{yeq}):
\bea
A_{11}(t)&=&\la y_1^2 \ra= k_B T \int_0^t dt'\int_0^t dt''~ G(t')~G(t'')~
\g(t'-t'')~, \nn \\
A_{22}(t)&=&\la y_2^2 \ra= k_B T \int_0^t dt'\int_0^t dt''~ \dot{G}(t')~
\dot{G}(t'')~ \g(t'-t'')~, \nn \\
A_{12}(t)&=&A_{21}(t)=\la y_1~y_2 \ra= k_B T \int_0^t dt'
\int_0^t dt''~ G(t')~\dot{G}(t'')~ \g(t'-t'')~.
\eea
Taking time-derivatives and after some manipulations one can get 
the following results for the variances:
\bea
\dot{A}_{11}&=&-2 k_B T~(~G~ \dot{G}+\f{F \dot{F}}{\omega^2}~)~, \nn \\
A_{11}&=& -k_B T ~(~G^2 +\f{F^2-1}{\omega^2}~)~, \nn \\
\dot{A}_{22}&=& -2k_B T ~(~\dot{G}~ \ddot{G} + \omega^2~ G~\dot{ G}~)~, \nn \\
A_{22}&=&-k_B T~ (~ \omega^2 G^2 + \dot{G}^2-1~)~, \nn \\
\dot{A}_{12}&=&k_B T~ (~ \f{d}{dt} (G~F)-\f{d}{dt}(G~\dot{G})~)~, \nn \\
A_{12}&=&k_B T ~(~G~F-G~\dot{G}~)~.\label{Aeqns}
\eea
These are of course identical to the results of Adelman since the
variances involve exactly the same Green's function. However the
mean values $\la x(t) \ra,~\la p(t) \ra$ are different and we will now
show that this leads to a different Fokker-Planck equation. 
The phase space probability distribution function for $\{x,~p\}$ is thus
\bea
P(x,p,t|x(0),p(0),0)=\f{1}{2 \pi (det[{\bf{A}}])^{1/2}} e^{-y_i A^{-1}_{ij}
  y_j/2}~. 
\eea
Following Adelman, let us define the following matrix:
\bea
 \T^{-1}(t)=\left( \begin{array}{cc} 
H(t) & G(t) \\ 
\dot{H}(t) & \dot{G}(t) \\
\end{array} \right)
 \eea
and new variables $\{ q_1,~q_2 \}$ through the transformation
\bea
\left( \begin{array}{c} q_1(t) \\ q_2(t) \end{array} \right) =  
\T(t) ~\left( \begin{array}{c} x_1(t) \\ x_2(t) \end{array} \right)~.   
\eea
Clearly we then have $q_1(0)=x(0),~q_2(0)=p(0)$.
In terms of these new variables the probability density $P$ takes the
following form:
\bea
\tilde{P}(q_1,q_2,t|x(0),p(0),0)&=&P(x_1,x_2,t|x(0),p(0),0) \nn \\ 
&=& \f{1}{2 \pi   (det[{\bf{A}}])^{1/2}} e^{-[q_i-q_i(0)] B^{-1}_{ij}
  [q_j-q_j(0)]/2}~, \\
{\rm where}~~~~{\bf{B}}&=& \T {\bf{A}} \T^T~. \nn 
\eea
Taking  time and space derivatives of $\tilde{P}$ gives the following relations:
\bea
\f{\p \tilde{P}}{\p t}&=&-\f{1}{2}~ \f{d}{dt} \ln{det[{\bf{A}}]}~ \tilde{P} -\f{1}{2}~
\dot{B}^{-1}_{lm}~[q_l-q_l(0)]~[q_m-q_m(0)] ~ \tilde{P}~,  \label{dwdt} \\
\f{1}{2} \f{\p^2 \tilde{P}}{\p q_l\p q_m} \dot{B}_{lm}&=&-\f{1}{2}
B^{-1}_{lm} \dot{B}_{lm} \tilde{P} +\f{1}{2}
B^{-1}_{li}~B^{-1}_{mj}~\dot{B}_{lm}~[q_i-q_i(0)]~[q_j-q_j(0)]  ~\tilde{P}
\nn \\
&=& -\f{1}{2}~ \f{d}{dt} \ln{det[{\bf{B}}]}~ \tilde{P} -\f{1}{2}~
\dot{B}^{-1}_{lm}~[q_l-q_l(0)]~[q_m-q_m(0)] ~ \tilde{P} ~, \label{dwdqq} 
\eea
where the last line follows from using the following identities for the 
symmetric matrix ${\bf{B}}$:
\bea
\f{d}{dt} det[{\bf{B}}] &=& \f{\p~ det[{\bf{B}}]}{\p B_{lm}}~ \dot{B}_{lm}= det[{\bf{B}}]~ B^{-1}_{lm}~
\dot{B}_{lm}~, \nn \\
\dot{{\bf{B}}}^{-1}&=&-{\bf{B}}^{-1}\dot{{\bf{B}}}{\bf{B}}^{-1}~. \nn 
\eea
We will also need the following result:
\bea
\f{\p \tilde{P}}{\p t}= \f{\p P}{\p t} + \dot{\T}^{-1}_{li} ~
{\T}_{im} x_m \f{\p P}{\p x_l}~\label{dpdt}~.
\eea
Using Eqs.~(\ref{dwdt}), (\ref{dwdqq}), (\ref{dpdt}) and transforming from
the variables $\{q_1~,q_2\}$ to $\{x_1,~x_2\}$ we finally get:
\bea
\f{\p P}{\p t} = -[~\dot{\T}^{-1} ~{\T}~]_{lm}~ x_m~ \f{\p P}{\p x_l}
-\f{d}{dt}\ln \Delta ~P+\f{1}{2}[~\T^{-1} \dot{{\bf{B}}} {\T^{-1}}^T~]_{lm}
\f{\p^2 P}{\p x_l \p x_m}~, \label{mast}
\eea
where $\Delta =det[\T^{-1}]=\dot{G}^2-G ~\ddot{G}$. We work out the explicit forms of
the various matrices and get:
\bea
\T^{-1}~\dot{\T}&=&-\dot{\T}^{-1}~\T~=~\left( \begin{array}{cc} 0 & -1 \\
\g_q & \g_p \end{array} \right)~, \\
\T^{-1} \dot{{\bf{B}}} {\T^{-1}}^T&=&\dot{{\bf{A}}} +\T^{-1}~ \dot{\T}~{\bf{A}} + (\T^{-1}~
\dot{\T}~{\bf{A}})^T \\ 
&=& \left( \begin{array}{cc} \dot{A}_{11}-2A_{12} &
  \dot{A}_{12}-A_{22}+\g_q ~A_{11} +\g_p~A_{12} \\
   \dot{A}_{12}-A_{22}+\g_q ~A_{11} +\g_p~A_{12} & \dot{A}_{22}+2 \g_q
   A_{12} +2 \g_p ~A_{22} \end{array} \right)~, \\
{\rm where~~~} \g_q&=&\f{\ddot{G}^2-\dot{G}~\dddot{G}}{\dot{G}^2-G
  ~\ddot{G}}~,  \\
\g_p&=&\f{G \dddot{G}-\dot{G}~\ddot{G}}{\dot{G}^2-G~
  ~\ddot{G}} =-\f{d}{dt}\ln {\Delta}~.
\eea
From Eq.~(\ref{Aeqns}) we get $\dot{A}_{11}=2 ~A_{12}$. Using this and 
expanding out Eq.~(\ref{mast}) gives:
\bea
\f{\p P(x,p,t|x(0),p(0),0)}{\p t} &=& -p \f{\p P}{\p x} + \g_q~x~\f{\p
  P}{\p p}+\g_p \f{\p (p P)}{\p p}+ D_p~ \f{\p^2P}{\p p^2} +
D_q~\f{\p^2 P}{\p x \p p}~,  \label{final} \\
{\rm where}~~~D_q &=&  \f{1}{2}\ddot{A}_{11}-A_{22}+\g_q ~A_{11}
+\f{\g_p}{2}~\dot{A}_{11}~,  \nn \\
D_p &=& \f{1}{2}\dot{A}_{22}+ \f{\g_q}{2} \dot{A}_{11} + \g_p
~A_{22}~. \nn
\eea
From Eq.~(\ref{gameq}) and Eq.~(\ref{kpeq}) it is clear that in the
classical limit:
\bea
K'(t) \to \f{k_B T}{\hbar} \g(t)~, 
\eea
which implies that the classical limits of the functions $K_q(t)$ and
$K_p(t)$ in Ref.~[\onlinecite{karrlein}] is:
\bea
\hbar~ K_q(t) \to A_{11},~~\hbar~K_p(t) \to A_{22}~.
\eea  
With this we immediately see that the coefficients of the equation for the Wigner
function, given by Eq.~($89$) in Ref.~[\onlinecite{karrlein}],
reduces in the classical limit, to those in the 
Fokker-Planck equation given by Eq.~(\ref{final}). 

\section{Correlated equilibrium initial conditions}
It is possible to understand the precise reason for the observed
correspondence between the Adelman results and the quantum
Fokker-Planck equation for the correlated initial conditions
considered in [\onlinecite{karrlein}]. We will now
show that the Langevin equation used by Adelman also follows 
from the same microscopic model if, instead of choosing a 
separable initial condition, we take a correlated initial state 
which is {\emph {prepared}} by making measurements on an equilibrium
state of the {\emph {coupled}} system and bath. We first make some 
comments on the physical significance of the special correlated initial
conditions that have been studied in the quantum case. The two initial
conditions correspond to (i) making measurements of the position
variable of the system at time $t=0$ on the equilibrium state, (ii) 
computing time dependent correlation functions of various system
variables, again in an equilibrium state. Especially it is to be  noted that the
measurements are made on the {\emph{equilibrium}} state of the
{\emph{fully coupled system and bath}}. 
Classically if we want to describe a similar situation, then the corresponding Langevin
equation  should decribe the time evolution of the system's phase
space variables $\{x,p\}$ {\emph{given}} that an initial measurement,
in the initial equilibrium state of the {\emph{coupled}} system and
bath, gave the values $\{x(0),p(0)\}$. 

Starting from the same model of reservoir with the full Hamiltonian given by
Eq.~(\ref{ham}) or equivalently Eq.~(\ref{eqham}) we again get an 
equation in the form of Eq.~(\ref{eqlang}) but the noise
correlations are not given by Eq.~(\ref{cnoise}). At $t=0$
the bath and system variables are correlated, hence $\la~ x(0)
~X_\a(0)~\ra \neq 0$ and the bath correlations are no longer give by
Eq.~(\ref{bathcor}) or its classical counterpart. A simplification now
occurs if we redefine the noise \cite{chaos} and write Eq.~(\ref{eqlang}) in the
following form:
\bea
 \ddot{x} &=& -\omega^2 x  - \int_0^t dt' \g (t-t') \dot{x}(t') +
\xi(t)~, \label{eqlang2} \\
{\rm where}~~~~ \xi(t) &=& \n(t) -\g(t) x(0) \nn \\
&=& \sum_\a c_\a \{~\cos (\omega_\a t)~ [~X_\a(0)-\f{c_\a}{\omega^2_\a}~x(0)~] +\f{\sin (\omega_\a t)}{\omega_\a}
\dot{X}_\a(0)~\}~. \label{noise2}
\eea
The statistical properties of the noise should be evaluated by
averaging over the bath variables. Following the discussion in the
previous paragraph it is clear that the distribution of the
bath variables is given by the {\emph{conditional}} probability of
$\{X_\a(0),P_\a(0)\}$ given that an initial measurement, in the
initial equilibrium state, on the system gave the values
$\{x(0),p(0)\}$.  
From the form of the Hamiltonian Eq.~(\ref{eqham}), we
get for the conditional distribution: 
\bea
Pr(\{X_\a(0),P_\a(0)\})&=& \f{e^{-\beta H_B}}{Z_B} \nn \\
{\rm where~~~~} H_B &=& \sum_{\a=1}^N [~\f{P_\a^2(0)}{2
  }+\f{\omega_\a^2 (X_\a(0)- \f{c_\a x(0)}{ \omega_\a^2}) ^2}{2}~]~,  \nn \\
Z_B &=& \int
\prod_\a dX_\a(0) dP_\a(0) e^{-\beta H_B}~. \nn
\eea
Averaging the bath variables over this distribution we then get:  
\bea
\la~[~X_\a(0)-\f{c_a}{\omega^2_\a}~x(0)~] \dot{X}_\a(0) ~\ra &=& 0~, \nn \\
\la~\omega_\a^2 [X_\a(0)-\f{c_a}{\omega^2_\a}~x(0)~]^2 ~\ra &=& k_B T~, \nn \\
\la~\dot{X}_\a^2 (0)~\ra &=& k_B T~.
\eea
Using these we then find that $\xi(t)$ is again a Gaussian stationary
process with exactly the same correlations 
as obtained for $\n(t)$ for separable initial
conditions. Thus
\bea
 \la \xi (t) \ra &=& 0~, \nn \\
 \la \xi (t) \xi(t') \ra &=& k_B T \g(t-t') \label{cnoise2}~.
\eea
The Langevin equation given by Eq.~(\ref{eqlang2}) with the noise given by
Eq.~(\ref{cnoise2}) is the starting equation of Adelman \cite{adelman}. 
Hence we have shown that the Langevin equation of Adelman is the
correct classical equation desribing a correlated initial state. 

\section{Discussion and Conclusions}
In this paper we have taken up the issue of the classical limit of the quantum
Fokker-Planck equation for a harmonic oscillator linearly coupled to a bath of
harmonic oscillators wherein the system and bath are initially uncorrelated.
Karrlein and Grabert \cite{karrlein} obtained the quantum Fokker-Planck
equation for some special correlated as well as separable initial conditions. Two kinds of 
correlated initial conditions were studied, the so called 
thermal initial condition \cite{hakim} and those describing the dynamics of equilibrium
correlation functions \cite{grabert}. On taking the classical limit of these equations
they found that they exactly reduced to Adelman's classical
Fokker-Planck equation \cite{adelman}. 
However for separable initial conditions they found that the classical
limit of their equation did not give Adelman's equation.
It was suggested that this indicated a  problem with separable initial conditions.

We have found a simple resolution to this somewhat intriguing point. 
We derived  the classical Langevin
equations of motion corresponding to the system of a single oscillator coupled  
to a bath of harmonic oscillators and with separable initial conditions. 
This equation differs from  the one used
by Adelman \cite{adelman} by an extra piece $-\g(t) x(0)$ which depends on initial conditions.
We then derived the Fokker-Planck equation for
this Langevin equation and found it to be the exact classical limit of the
quantum equation of Karrlein and Grabert \cite{karrlein} for separable initial
conditions. We have also shown why for the case of correlated 
initial conditions the Wigner function equation
of the reduced system dynamics in the quantum
regime \cite{karrlein} reduces to the Adelman  
equation in the classical limit. 
We have shown that the Langevin equation used by  Adelman
corresponds to the situation where the initial state is the
equilibrium state of the coupled system and reservoir. 
Thus the basic point is that {\emph{correlated and uncorrelated initial
conditions lead to different classical Langevin equations}} and hence to
different Fokker-Planck equations. 
In conclusion we have shown  that for the case of a single harmonic 
oscillator coupled to a bath of harmonic oscillators
there is an exact correspondence
between the classical Fokker-Planck equation and the quantum equations
for the Wigner function both for some special correlated as well as
uncorrelated initial conditions.

\end{document}